# Radio Frequency Interference Management with Free-Space Optical Communication and Photonic Signal Processing

**Yang Qi and Ben Wu\***
*Department of Electrical and Computer Engineering, Rowan University, Glassboro, NJ 08028, USA.*
*wub@rowan.edu*

**Abstract:** We design and experimentally demonstrate a radio frequency interference management system with free-space optical communication and photonic signal processing. The system provides real-time interference cancellation in 6 GHz wide bandwidth.

## 1. Introduction

Current communication techniques evolve as the need for channel capacity increases. As spectrum expands from 2.4 GHz for 4G LTE networks to sub-6 GHz for 5G networks, higher frequency bands are introduced for industrial and business use to provide faster transmission and more band capacity [1,2]. However, it is neither realistic nor cost efficient to upgrade the current communication systems to higher frequency bands all at once, hybrid systems that include both lower and higher frequency bands are needed. When these communication systems operate side by side, interference appears between bands. Interference management systems that can offer a wide bandwidth for different communication protocols and fast procession time are needed.

The proposed system uses photonic signal processing and free-space optical communication methods for interference management. Photonic signal processing uses optical modulators to modulate RF signals to optical frequencies, since photonic devices can process signals in THz bandwidth, it can achieve wideband signal processing in real-time for current commercial radio frequency (RF) bandwidth or even higher spectrum. This method has been proven to work on spectral filtering [3,4] and self-interference cancellation [5]. Free-space optical communication can provide precise directional communication for stealth communication.

In this project, we design and experimentally demonstrate a radio frequency interference management system that achieves interference cancellation in 6 GHz bandwidth. When multiple RF transmitters and receivers operate in the same bandwidth simultaneously, one signal of interest (SOI) from one pair of transmitters and receivers can interfere with signals from other pairs, those signals become the interference to the designated receiver. To cancel the interference, an interference reference signal is needed to send to the designated receiver, this reference signal is modulated on the optical frequency and sent through a free-space optical channel in this system, which does not occupy an extra RF spectrum. Two pairs of transmitter-receiver are tested as a model for multiple users, SOI sent through the designated pair are modulated onto optical frequency by the photonic signal processing method and mix with the interference reference signal, the reference signal cancels the interference and recovers the SOI.

## 2. Experiment Setup

Fig. 1. shows the experimental setup of the Radio Frequency Interference Management with Free-Space Optical Communication and Photonic Signal Processing with two transmitter-receiver pairs. The SOI is sent through Transmitter A and received by Receiver A, the interference is generated as noise and sent through Transmitter B, which can also be received by Receiver A. At the point of Receiver A, a mixed RF signal of SOI and interference are modulated on an optical intensity modulator at the laser frequency of 1544nm. The interference signal is split as the reference signal, modulated at the laser frequency of 1560nm, and sent through a pair of optical collimators as the optical free-space channel. The laser frequencies for the designated receiver and reference signal are set differently to avoid optical self-interference. The modulated mixture of SOI and interference combine with the reference signal at the optical fiber combiner, the phase, and intensity difference can be adjusted by tuning the bias voltage on the optical intensity modulators, adjusting optical attenuators, and setting tunable delays. Once the reference signal is phase-matched and intensity-inverted with the interference signal, they will cancel each other and only the SOI will remain to be transformed to radio frequency by a photodiode (PD) and analyzed by Vector Signal Analyzer.

The system tested as shown has two transmitter-receiver pairs, more pairs can be introduced by adding a third optical free-space channel. By changing the bias voltage, attenuations, and delays for each channel, the weight for each reference signal changes before they combine and cancel certain elements of the interference to recover the SOI. The system tested uses Keysight M9381A PXIe Vector Signal Generator to generate an LTE FDD signal with 64QAM modulation at the carrier frequency of 2.4 GHz. The interference signal is generated by Hewlett Packard ESG-D4000A as frequency modulation of Gaussian noise at the carrier frequency of 2.4 GHz. The received RF signal is sent and analyzed by Keysight M9391A PXIe Vector Signal Analyzer for spectrum observation, constellation reconstruction, and error vector magnitude (EVM) calculation.

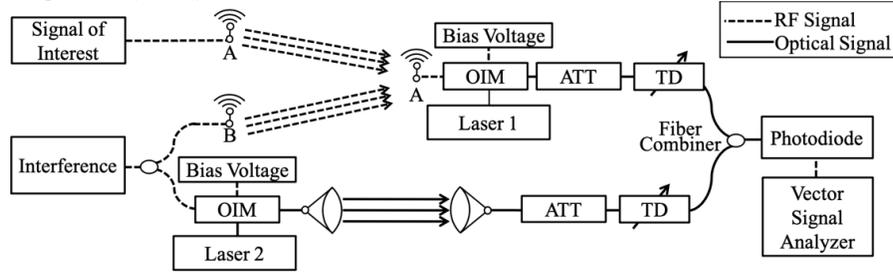

Fig. 1. Experimental setup. (A, B: RF antenna, OIM: optical intensity modulator, ATT: optical attenuator, TD: optical tunable delay)

## 3. Results and Analysis

Fig. 2. shows the experimental results for Radio Frequency Interference Management with Free-Space Optical Communication and Photonic Signal Processing. Fig. 2. (a) shows the spectrum of the received signal with and without photonic interference management. The black line shows the spectrum of the received RF signal, a mixture of a 40 MHz bandwidth interference signal and a 5 MHz LTE FDD 64QAM signal both at a carrier frequency of 2.4 GHz, the red line shows the spectrum with the photonic interference management, where the SOI is separated. Fig. 2. (b) (c) shows the constellation diagrams for received signal without (b) and with (c) photonic interference management. The constellation diagram improves after the interference management and EVM improves from 33.876% to 4.171%.

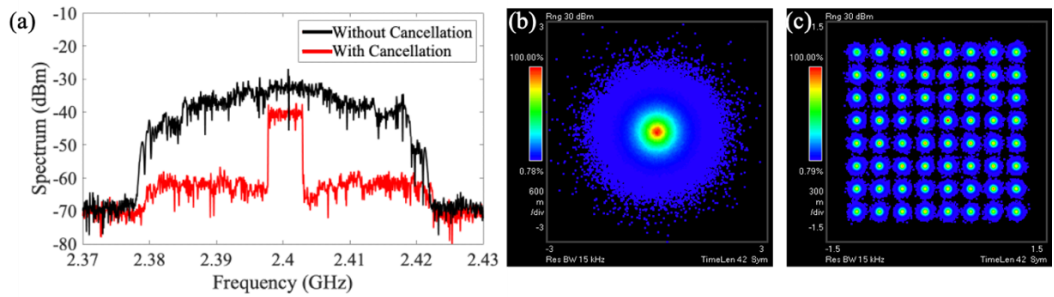

Fig. 2. (a) Spectrum with and without photonic interference management. Constellation diagrams for received RF signal without (b) and with (c) photonic interference management.

## 4. Conclusion

We design and experimentally demonstrate the Radio Frequency Interference Management with Free-Space Optical Communication and Photonic Signal Processing. The system sends a reference signal through optical free space to the receiver which combines with the received mixed RF signal to separate SOI. The system is tested work with frequency ranges up to 6 GHz, so a wide bandwidth of different communication protocols can be utilized, without occupying extra RF bands. By testing with SOI of 64QAM, EVM drops from 33.876% to 4.171%, different modulation methods including QPSK, 16QAM, 64QAM, and 256QAM are tested work to recover a low error vector magnitude. Two transmitter-receiver pairs are tested in the experiment, more transmitter-receiver pairs can be added by adding a third optical free-space channel to achieve multiuser management.

## References

[1] I. Akyildiz, D. Dutierrez-Estevez and E. Reyes, "The evolution to 4G cellular systems: LTE-Advanced" in Physical Communication, vol. 3, pp. 217-244.
[2] M. Shafi, A. Molisch, P. Smith, T. Haustein, P. Zhu, P. Silva, F. Tufvesson, A. Benjebbour and G. Wunder, "5G: A Tutorial Overview of Standards, Trials, Challenges, Deployment, and Practice," in IEEE Journal on Selected Areas in Communications, vol. 35, no. 6, pp. 1201-1221.
[3] Q. Liu and M. P. Fok, "Dual-function Frequency and Doppler Shift Measurement System Using a Phase Modulator Incorporated Lyot Filter," in Optical Fiber Communication Conference, paper Th2A.36.
[4] Q. Liu, J. Ge and M. P. Fok, "Microwave photonic multiband filter with independently tunable passband spectral properties," in Optics Letters, vol. 43, pp. 5685-5688.
[5] Y. Qi, and B. Wu, "Photonic Interference Cancellation with Free-Space Optical Link for Cellular Networks," in Frontiers in Optics / Laser Science, paper FTh5E.2.